\documentclass[aps,prl,twocolumn,showpacs,floatfix]{revtex4}

\usepackage{hyperref}
\usepackage{amsmath,bbm,color}

\definecolor{grey}{rgb}{0.75,0.75,0.75}
\definecolor{orange}{rgb}{1.0,0.5,0.5}
\definecolor{brown}{rgb}{0.5,0.25,0.0}
\definecolor{pink}{rgb}{1.0,0.5,0.5}
\usepackage{amsbsy,latexsym}
\usepackage{amsfonts}
\usepackage{amssymb}
\usepackage[mathscr]{eucal}
\usepackage{graphics}

\usepackage{amsbsy,latexsym}
\usepackage{amsfonts}
\usepackage{amssymb}
\usepackage[mathscr]{eucal}
\usepackage{graphics}

\newcommand{\tr}{{\rm tr}\,}

\newcommand{\ket}[1]{\vert #1 \rangle}
\newcommand{\bra}[1]{\langle #1 \vert}

 \newcommand{\defi}{\stackrel{\mbox{\tiny def}}{=}}
 \newcommand{\ketbrad}[1]{|#1\rangle\!\langle #1|}
 \newcommand{\ketbra}[2]{|#1\rangle\!\langle #2|}

\begin{document}

\title{Recycling of quantum information: Multiple observations of quantum systems}

\author{P. Rap\v{c}an$^1$, J.~Calsamiglia$^2$, R.~Mu\~{n}oz-Tapia$^2$, E.~Bagan$^{2,3}$ and V. Bu\v{z}ek$^{1}$}

\affiliation{$^1$ Research Centre for Quantum Information, Slovak Academy of Sciences, D\'{u}bravsk\'{a}
cesta 9, Bratislava, Slovakia
\\
$^2$Grup de F\'{\i}sica Te\`orica \& IFAE, Edifici Cn, Univ. Aut\`onoma de Barcelona, 08193 Bellaterra (Barcelona) Spain
\\
$^3$Department  of Physics and Astronomy, University of New Mexico, Albuquerque , NM 87131, USA
}

\date{8 August 2007}

\begin{abstract}
 Given a finite number of copies of an unknown qubit state
 that have already been measured optimally, can one still
  extract any information about the original unknown state?
  We give a positive answer to this question and quantify the information obtainable
 by a given observer as a function of the number of copies in the ensemble, and of the number of independent observers that, one after the other, have independently measured the \emph{same} ensemble of qubits before him.
 The optimality of the protocol is proven and extensions to other states and encodings are also studied. According to the general lore, the state after a measurement has no information about the state before the measurement. Our
results manifestly show that this statement has to be taken with a grain of salt, specially in situations where the quantum states encode confidential information.


%

\end{abstract}


\maketitle

One of the major questions in the interpretation of  quantum mechanics is to assert the reality of the wave function or the quantum state.
Despite the opinion galore, which also includes rejecting the necessity of attributing reality to a quantum state \cite{fuchs_quantum_2000}, there is consensus in that all information on the state of a system is contained in the wave function (in the sense that it provides the right outcome probabilities for each conceivable measurement on the system). Since all this information is not accessible by a single measurement and, on top of that, quantum formalism only gives outcome probabilities, the meaning of wave function has been traditionally associated to an  {\em infinite} ensemble of identically prepared quantum systems (something which cannot be taken literally, but only as a conceptual notion). Ground-breaking experiments with individual quantum systems (see e.g. \cite{Haroche,Blatt}) and the advent of quantum information technology have
brought the focus to individual systems, away from the infinite ensemble picture.

The seminal works of Helstrom \cite{Helstrom} and Holevo \cite{Holevo} have provided the means to quantify the amount
of information that can be obtained from measurements on finite-size
ensembles of quantum systems. In particular, we can now compute
limits on the amount of classical information about the original
state that can be obtained when we measure a single system. However,
any gain of information about such state is
accompanied by a disturbance of the measured system (see e.g.
\cite{Fuchs,Banaszek}).  Once the quantum system has been optimally measured one cannot extract {\em more} information from it any more; i.e. an experimentalist who has performed an optimal measurement cannot learn more about the state by performing further measurements on the same system.
For this reason it is often believed that the state after a
projective measurement does not contain any information about the
original state, only the classical information gathered by the
experimentalist can be transmitted to other observers.  A more
careful analysis, however, shows that \emph{there is} some information left in the posterior state. Specifically, let us assume that the experimentalist does not share
her classical information (about her measurement device or about the obtained outcomes) with any other observer. Is it still possible for a new observer to obtain information about the original preparation of the system?
And, if so, how does this information degrade through a sequence of independent measurements performed by non-communicating observers? Can posterior states left over after running a quantum communication
protocol compromise its security?

The problem presented here also touches upon another thorny problem in quantum mechanics, namely, determining its range of validity or the so called quantum to classical transition  \cite{Zurek1991}. Is quantum mechanics restricted to small scale physics?  The microscopic world is governed by the rules of quantum mechanics, which are in sharp contrast with the rules of classical physics that govern the macroscopic world.
Can one explain this very different behavior within quantum mechanics in a consistent fashion? Before attempting to answer these questions it is important to recognize what are the essential features that appear to be so different in the classical and quantum worlds.  The problem at hand sheds some light on one of these differential aspects, which is the fragility of the information encoded in quantum states versus the recyclability of classical information. Indeed the information encoded in a classical system can be accessed by an unlimited amount of (careful) observers without degrading while, as we shall discuss here, quantum mechanics allows to recycle some amount of information but it degrades with the number of observers.

In this Letter we present a quantitative investigation of the amount
of information that can be extracted from a quantum
system by a series of sequential observers  who are not allowed to communicate.
In particular we show that this information, quantified by the fidelity (see below), decays exponentially with the number of observers.
We show that the larger the ensemble of identically prepared
systems is, the more observers can gain a {\em sizeable} amount of the
classical information encoded in the original state. In other words, the information encoded in large ensembles
of quantum systems behaves ``classically'', in the sense that  it is robust with
respect to observations. If we relax the condition of having identically prepared
copies, there are even more robust ways to encode
classical information. We present the optimal one for quantum systems
made out of $N$ spins. We also discuss briefly other extensions of
the problem, such as the optimal equitable distribution of the encoded
information among the observers. In this case,
each observer in the sequence has to have the same amount of information.

We consider quantum systems carrying some classical information,
i.e., values of some variables which we encode in the quantum states of the systems. Any observer who wishes to access (estimate) this information must
perform a measurement and, knowing the encoding, interpret its
outcomes. The estimation will depend on the prior knowledge about
the encoded information which, given the encoding, induces a prior
knowledge about the encoding state.  We will address the problem from a Bayesian point of view.
We thus need a figure of
merit to quantify the accuracy of the estimation. In this approach, an optimal
measurement is one that maximizes the average figure of merit with
respect to the prior and to all possible outcomes. Here, since the
system will be measured by a series of observers, we additionally
require realizations of the measurements that produce the minimal
disturbance to the state.


Let us consider as a specific example of classical information the
direction of a given unit vector $\vec{n} (\theta, \phi)$ in three-dimensional space, and
further restrict ourselves to encoding quantum systems made out of
$N$ spin-$\frac{1}{2}$ particles. We assume that none of the observers has any
knowledge whatsoever about the encoded direction. In this case the  prior
probability distribution of $\vec n(\theta,\phi)$ is uniform on the 2-sphere ${\mathbb S}^2$ and is thus given by
 $dn=\sin\theta d\theta d\phi /(4\pi)$.

 Assume now that $k$ observers estimate $\vec n(\theta,\phi)$
in succession by using identical and optimal measurement devices.
Note however that the observers are assumed not to communicate. In particular, they do not know about the relative
orientation of their measurement devices  (e.g.,  of their Stern-Gerlachs, in the case of a single spin-$\frac{1}{2}$ particle). We may also think of them as observers whose
reference frames are chosen randomly relative to each other. Further, they are fully ignorant about each other's outcomes.
%
%
%

Following Refs.~\cite{Massar1995,Derka1998,Gisin1999,Bagan2000}, we here use the
single-qubit fidelity, $f (\vec{m}_k, \vec{n}) = \frac{1}{2} (1 +
\vec{m}_k \cdot\vec{n})$ as the figure of merit. The $k$'th observer's
success in gaining knowledge about $\vec n(\theta,\phi)$ is given by the
mean of the  fidelity, $F_k$, over all encoded and estimated
directions
\begin{equation}
  F_k  =  \int dn 
\sum_{\vec m_k}  
\frac{1 +
\vec{m}_k \cdot\vec{n}}{2} p(\vec{m}_k| \vec{n})\defi\frac{1+\Delta_k}{2}
\, , \label{eq:AvgCost}
\end{equation}
where  $p (\vec{m}_k| \vec{n})$ is the conditional probability of
obtaining an estimate $\vec{m}_k$ if the signal state was encoding
the direction~$\vec{n}$, and the sum can run over a discrete as well as a
continuous set of estimates $\{\vec{m}_k\}$, each of them inferred from an outcome of the $k$th observer's measurement.

Let us first consider a  single qubit ($N=1$), which is sufficient
to illustrate our main points. In this case the Bloch vector of a
spin-1/2 pure state intrinsically encodes the direction $\vec{n}(\theta,\phi)$, so we can
write the signal states as $\rho_0(\vec n)=\ket{\vec n}\bra{\vec n}$. The
optimal estimation protocol is well known {\cite{Massar1995}}. The
observers measure the spin component along {\em any}  direction $\vec m_1$, via a
projective (Stern-Gerlach-like) measurement, make an estimate that
corresponds to the direction of the outcome $x=\pm$, i.e., $\vec m_1(x)=x\,\vec m_1$, and pass the posterior state $\rho_1$ to the next observer. According to her point of view (the previous measurement axis $\vec m_1$ is a uniformly-distributed random unit vector), we may consider the state $\rho_1$ to be
\begin{eqnarray}
\rho_1&\!\!\!=\!\!\!\!&\int \!\!dm_1 \sum_x 
\tr\!\!\left[\ket{\vec m_1(x)}\!\bra{\vec m_1(x)}\,\rho_0(\vec n)\right]
\ket{\vec m_1(x)}\!\bra{\vec m_1(x)}\nonumber\\
&\!\!\!\!=\!\!\!\!&
\int \!\!dm_1\, \tr\!\!\left[2\ket{\vec m_1}\bra{\vec m_1}\rho_0(\vec n)\right]\; \ket{\vec m_1}\!\bra{\vec m_1}.
\label{eq:S-G=cov}
\end{eqnarray}
The second equality in (\ref{eq:S-G=cov}) shows that one gets the same state $\rho_1$ regardless whether a Stern-Gerlach-like or a continuous covariant POVM measurement [$O(\vec m)=2\ket{\vec m}\bra{\vec m}$, which implies that $\int dm \,O(\vec m)=\openone$] is used.  One can show that this is a general result that applies to any projective measurement, and that it will also hold in the extensions we consider below. Covariant POVMs are known to be optimal in state estimation too \cite{Holevo}. So, for the sake of simplicity and with no loss of generality, we will mainly be concerned with continuous covariant POVM throughout the rest of the letter.
%
For the first observer we have 
\begin{eqnarray}
    \Delta_1 \!\!=\!\! \int dn\int dm_1\, \vec n\cdot\vec m_1\, \tr\!\!\left[O(\vec m_1)\rho_0(\vec n)\right] =\frac{1}{3}
    \label{eq:Delta1}
\end{eqnarray}
which is a well known result.

At this point in the discussion, it should be clear that the crux of the matter is knowing the right
description of the state that is passed to the second
(and subsequent) observer(s). This is
determined by the specific realization of the measurement, i.e., by the
precise Kraus decomposition of the POVM elements
$O(\vec{m}_k)=A^{\dag}(\vec{m}_k) A(\vec{m}_k)$.
In the case of Eq.~(\ref{eq:S-G=cov}), for instance, by simply rearranging bras and kets, we see that
\begin{equation}
\rho_1=\int dm_1\left( \sqrt2\ket{\vec m_1}\!\bra{\vec m_1}\right)^\dagger \rho_0(\vec n)\left( \sqrt2\ket{\vec m_1}\!\bra{\vec m_1}\right),
\label{eq: channel01}
\end{equation}
i.e., $A(\vec m_1)= \sqrt2\ket{\vec m_1}\!\bra{\vec m_1}$, which is obviously the optimal choice of the Kraus operator.
It should also be clear that the worst possible choice is $A(\vec{m}_k)=\sqrt2 \ket{-\vec{m}_k}\!\bra{\vec{m}_k}$, in which case the state passed to the second observer points in the direction opposite to the first observers guess.
%

Proceeding along these lines, after $k$ measurements we have
\begin{equation}\label{eq:rhok}
   \rho_k=\int  \ket{\vec m_k}\!\bra{\vec m_k} \prod_{j=1}^k dm_j 
\tr\left[O(\vec m_j)\rho_0(\vec m_{j-1})\right] ,
\end{equation}
where we have defined $\vec m_0\defi\vec n$. Similarly, the fidelity after
measurement $k$  is obtained from
\begin{equation}\label{eq:Deltak}
    \Delta_k=\int  dn\,\vec n\cdot\vec m_k  \prod_{j=1}^k dm_j
\tr\left[O(\vec m_j)\rho_0(\vec m_{j-1})\right].
\end{equation}
Integrating over $\vec n$ in this last equation, we have
\begin{equation}
\Delta_k=\Delta_1 \!\!\int \!\! dm_1\,\vec m_1\cdot\vec m_k \prod_{j=2}^k
dm_j \tr\left[O(\vec m_j)\rho_0(\vec m_{j-1})\right] ,
\end{equation}
where we have used that $ \int  \! dn\, \vec n\,  \tr\left[O(\vec m_1)\rho_0(\vec n)\right]
=\Delta_1 \vec m_1 $. Similarly, integrating over $\vec m_1$ we have,
\begin{equation}
\Delta_k=\Delta_1^2 \!\!\int \!\! dm_1\,\vec m_1\cdot\vec m_k \prod_{j=3}^k
dm_j \tr\left[O(\vec m_j)\rho_0(\vec m_{j-1})\right] .
\end{equation}
This process can be iterated to give $\Delta_k=\Delta_1^k$. Hence,
\begin{equation}
F_k=\frac{1+\Delta_1^k}{2}=\frac{1}{2}\left( 1+\frac{1}{3^k}\right).
 \label{eq:Fk}
\end{equation}

Eq.~\eqref{eq:Fk} gives the maximum mean fidelity that the
$k$th observer  can achieve. We see that all
the observers can gain some information about the original direction $\vec n(\theta,\phi)$,
but the fidelity of their estimates degrades {\em exponentially} with their
tally number within the sequence of observers.

We can also view the measurement process as a channel
$\rho_{i+1}=\mathcal{L}(\rho_{i})$. This is clearly the case in Eq.~(\ref{eq: channel01}), from which one obtains $\mathcal{L}(\rho_{i})=\rho_i/3+\openone/3$. Hence, $\cal L$
corresponds to a depolarizing channel with shrinking factor~$\eta=1/3$. It is not difficult to obtain the expression of~${\cal L}$ for a generalized measurement with Kraus
operators~$O(\vec{m}_i)=A^{\dag}(\vec{m}_i)A(\vec{m}_i)$. It is
\begin{equation}\label{channel}
    \mathcal{L}(\rho_{i})= \eta \rho_i+
    (1-\eta)\frac{\openone}{2}=
    \frac{c-1}{3}\rho_i+\frac{4-c}{3}\frac{\openone}{2} ,
\end{equation}
where $c=\int dm_i |\tr A(\vec{m}_i)|^2$. Note that the constant $c$
quantifies the disturbance inflicted to the state by the particular realization $\{A(\vec m_i)\}$ of
the measurement~\cite{Banaszek2001}. Its minimum value,  which gives  the
maximal disturbance, is $c=0$ and corresponds to a realization
where the Kraus operators are traceless. The choice $A(\vec{m}_k)=\sqrt2 \ket{-\vec{m}_k}\!\bra{\vec{m}_k}$ discussed above provides a good example of this situation. The maximal
value of $c$, which produces the minimal disturbance, is given by the dimension of the Hilbert
space: $c=2$. This value is obtained with~$A(\vec{m}_k)=\sqrt2
\ket{\vec{m}_k}\!\bra{\vec{m}_k}$ and yields the optimal shrinking
factor $\eta=1/3$. It is now easy to reobtain the fidelity~$F_k$, since it will correspond to the product of shrinking
factors, i.e., $\Delta_k=\eta^k$. Therefore, the maximal fidelity is that given in~(\ref{eq:Fk}).

Let us now tackle a more involved problem. That of encoding $\vec n(\theta, \phi)$ in a state made out of
$N$ spin-$\frac{1}{2}$ systems. We consider the covariant encoding given by
%
  $  \rho_0(\vec{n})=U(\vec{n})\rho_0 U^\dag(\vec{n})$,
%
where $\rho_0$ is a fiducial state pointing along a fixed direction,
say $\vec{z}$, i.e., $[J_z,\rho_0]=0$, where~$J_z$ is the projection of the total spin of the system of size $N$ along $\vec z$. By $U(\vec{n})$ we denote  the unitary
representation on $({\mathbb C}^2)^{\otimes N}$ of the rotation that takes~$\vec{z}$ into~$\vec{n}$. As for a single spin-$\frac{1}{2}$ system,
we can stick to continuous POVMs and recall that optimality implies  that the POVM elements must be of rank one
$\{O(\vec m_k)=\ketbrad{\Psi_{\vec m_k}}\}$. In particular this means that we can interpret the associated Kraus operators
$A_{\vec m_k}=\ketbra{\Phi_{\vec m_k}}{\Psi_{\vec m_k}}$ as the ones corresponding to a measure and prepare channel, where
after projecting on  $\ket{\Psi_{\vec m_k}}$ the state is prepared in state  $\Omega(\vec m_k)=\ketbrad{\Phi_{\vec m_k}}$.
Following along the same lines as in the single copy case we have that the state after $k$ measurements is
\begin{equation}\label{eq:rhok-general}
    \rho_k(\vec n)=\!\!\int \!\! \Omega(\vec m_k)\! \prod_{j=1}^k dm_j \,
    \tr\left[\Omega(\vec m_{j-1})O(\vec m_j)\right],
\end{equation}
where we have defined 
$\Omega(\vec
m_0)\defi\rho_0(\vec n)$. Similarly, the fidelity is determined by
\begin{equation}\label{eq:Deltak-general-1}
    \Delta_k=\!\!\int \! dn\,\vec n\cdot\vec m_k
    \prod_{j=1}^k dm_j \,\tr\left[\Omega(\vec m_{j-1})O(\vec m_j)\right].
\end{equation}
We can perform the integration in \eqref{eq:Deltak-general-1} by
noticing the recursion relation
$
    \Delta_k=\Delta_{k-1} \tilde{\Delta}[\Omega(\vec m_{k-1})]
$,
where $\{1+\tilde{\Delta}[\Omega(\vec m_{k-1})]\}/2$ is the fidelity with which the $(k-1)$'th observer would estimate the direction $\vec n$ by performing her measurements on $\Omega(\vec n)$. Iterating this relation we immediately obtain the  final compact result
\begin{equation}\label{eq:Deltak-general-2}
    \Delta_k=\prod_{j=0}^{k-1}\tilde\Delta(\Omega_j).
\end{equation}
%

We can now address some specific situations within this $N$ spin-$\frac{1}{2}$ system scenario.  Let us first consider an encoding state consisting of identical $N$ copies of $\ket{\vec n}$, which we may refer to as parallel spin case. The initial state has the form $\rho_0^{\rm par}(\vec
n)=\ket{\vec n}\!\bra{\vec n}^{\otimes N}$. A
natural choice of measurement is given by the Kraus operator of the same form, $A^{\rm par}(\vec m_{j})=\sqrt{N+1}\ket{\vec m_{j}}\!\bra{\vec m_{j}}^{\otimes N}$ for all observers.
This implies that   $ \Omega^{\rm par}(\vec{m}_j)= \ket{\vec m_j}\!\bra{\vec m_j}^{\otimes
    N}
$
for the state passed to observer~$j+1$.
One obtains~\cite{Massar1995} $\tilde\Delta[\Omega^{\rm par}(\vec{m}_j)]=N/(N+2)$,
and
\begin{equation}\label{eq:Fk-parallel}
    F^{\rm
    par}_k=\frac{1}{2}\left[1+\left(\frac{N}{N+2}\right)^k\right].
\end{equation}
With increasing number of qubits, the
observers can estimate the state of the system more reliably. More
interestingly, even though the first observer takes advantage of the
size of the quantum system to extract all the information she can, still, with increasing number of qubits, more independent observers can infer the classical
information about  the initial preparation reliably.
To be more specific, we see that if number of qubits grows as $N\sim k^\alpha$ then for  $\alpha>1$ we recover the classical behavior in the sense that a large number of observers can infer the original direction with reasonable precision, i.e. $F^{\rm
    par}_k\to 1$ as $k\to \infty$.

However, it turns out that the above choice of Kraus operators  is not
optimal (nor is the initial encoding state). As follows from Eq.(\ref{eq:Deltak-general-2}),
the optimal choice must maximize $\tilde\Delta(\Omega_j)$ for each observer. This is accomplished by
%
$\Omega_i(\vec m_i)=U(\vec m_i)\,\ket \Phi\!\bra \Phi\, U^\dagger(\vec
m_i);\quad i\ge 0.$
%
In the standard bases of the $SU(2)$ irreducible representations, $\{\ket{J,M}\}_{M=-J}^J$, $J=0,1,\dots,N/2$
(for simplicity we assume that $N$ is even), the state $\ket\Phi$ has the form
%
$
\ket{\Phi}=\sum_J \Phi_J \ket{J,0}
$.
 %
The (real) components $\Phi_J$ can be arranged as a vector $\mbox{{\boldmath $\Phi$}}=(\Phi_0,\dots,\Phi_{N/2})^t$
and are defined by the following properties: (a)
 $\mbox{{\boldmath $\Phi$}}^t\mbox{{\boldmath $\Phi$}}=1$ (normalization); (b)
define the symmetric matrix $\mbox{{\boldmath $M$}}=[m_{ij}+m_{ji}]$, where $
m_{jl}=j(4j^2-1)^{-1/2} \delta_{j+1,l}
$. Then $\mbox{{\boldmath $\Phi$}}$ is the eigenvector of $\mbox{{\boldmath $M$}}$ with largest eigenvalue~\cite{Bagan2000,Bagan2001}.
The optimal measurement is realized by the Kraus operators
%
$A^{\rm op}(\vec m_{i})=U(\vec m_i)\,\ket \Phi\bra \Psi\, U^\dagger(\vec m_i); \quad i\ge1 \, ,
$
%
where $ \ket{\Psi}=\sum_{J=0}^{N/2} \sqrt{2J+1}\ket{J,0}$. 
In this case
%
$\tilde\Delta(\Omega_i^{\rm op})=x_{N/2+1}$ ,
%
where $x_{N/2+1}$ is the largest zero of the Legendre polynomial
$P_{N/2+1}(x)$. Thus
$
F^{\rm op}_k=[1+(x_{N/2+1})^{k}]/ 2.
$

Of course had we started with the  parallel spins instead of the optimal state, we would only have to change the measurement to $A(\vec m)=U(\vec m)\ket \Phi({\bra{\vec m}})^{\otimes N}$ for the first observer and accordingly obtain the optimal value for parallel spins $\Delta^{\rm par}_{k}=N/(N+2) (x_{N/2+1})^{k-1}$.

It is known that the largest zero of the Legendre polynomial $P_n(x)$ goes as $ x_n=1-\xi_0^2/
(2n^2)+\cdots$ for large $n$, where $\xi_0\approx2.4$ is the first zero of the Bessel
function~$J_0(x)$~\cite{Bagan2001}. Hence, asymptotically
\begin{equation}\label{eq:optimal-N-fideltiy-asympt}
F^{\rm op}_k=\frac{1}{2}\left[
1+\left(1-\frac{2\xi_0^2} {N^2}\right)^{k} \right] ,
\end{equation}
which is the maximum ``recycled" fidelity that can be obtained for
large $N$ spin systems. In this case we see  that the size $N$ for a system of spins to be considered ``classical" is quadratically smaller
as compared to
the parallel case, i.e. $F^{\rm op}_k\to 1$ if $N\sim k^\alpha$ with $\alpha>1/2$ (instead of $\alpha>1$, above).
The entanglement of the encoding $\ket{\Phi}$ is
behind this improvement. Entanglement makes the quantum system less
fragile upon sequential observations.

One might consider a slightly different scenario  where
it is required that all $k$ independent observers gain the
same and maximal information about the initially encoded direction. This would
be an instance of ``redistribution'' of information as one has for example in the universal cloning machine. In this case it is clear
that the observers have to do weak measurements. Only the last observer can do a fully sharp measurement.
We find that the optimal protocol yields a fidelity that for large
$k$ degrades as the square root of the number of observers
$\Delta_{k}\sim 1/\sqrt{k}$ \cite{footnote}.
The proof  involves Kraus operators of full rank and cannot be treated  as a simple measure and prepare channel.
The same asymptotic behavior $\Delta_{k}\sim 1/\sqrt{k}$ is recovered if we are asked to devise a protocol where the last $k$'th
observer obtains the maximal fidelity after $k$ identical measurements. Finally, most of our results can be extended
rather directly to qudits. All these issues will be treated elsewhere.

%

This work was supported by the European Union projects QAP, CONQUEST, by projects
CE-PI, APVT-99-012304, VEGA, by the European Social Fund via the project IAS,
by the Spanish MEC contracts FIS2005-01369, QOIT (Consolider-Ingenio 2010) and by the Catalan government, CIRIT SGR- 00185.
JC and EB acknowledge MEC financial support through the Ram\'on y Cajal program and travel grant PR2007-0204, respectively. EB, RMT and JC also
thank the Benasque Center for Sciences for providing an inspiring research atmosphere.


\begin{thebibliography}{1}
\bibitem{footnote}
This decays slower than for the $1\to N$ universal cloning machine that decays as $1/k$.
Of course there is no contradiction here, since the asymptotic value of the fidelity is $1/2$ for recycling while it reaches
$2/3$ in the universal cloner/one copy state estimation.
\end{thebibliography}
\end{document}